\documentstyle[preprint,aps,tighten,epsf,floats]{revtex}
\newcommand{\ben}{\begin{displaymath}}
\newcommand{\een}{\end{displaymath}}
\newcommand{\be}{\begin{equation}}
\newcommand{\ee}{\end{equation}}
\newcommand{\bea}{\begin{eqnarray}}
\newcommand{\eea}{\end{eqnarray}}

\begin{document}
\draft
\title{
How the relation $e^{2}/4\pi\approx 1/137$ may be obtained in the framework of Quantum Electrodynamics}
\author{ G.~S.~Japaridze}
\address{Center for Theoretical Studies of Physical Systems, Clark Atlanta
University, Atlanta, GA 30314}
\maketitle

\begin{abstract}
In this pedagogical note it is demonstrated how the numeric value of fine structure constant may be established by comparing results following from the calculations in the framework of Quantum Electrodynamics with the experimental data. As an observable, the coefficient of $1/r$ in the Coulomb law is used. 
\end{abstract}
\newpage 
\subsection*{Introduction}
It is a wide-spread perception that Quantum Field Theory \cite{bjerken}, the  framework for operating with the fundamental degrees of freedom, serves as the basis of of our understanding of the Universe. According to Quantum field Theory a full information about the physical system under consideration is provided by the Lagrangian of the corresponding field-theoretical model. The only unknowns in this approach are the values of the parameters appearing in the Lagrangian, e.g. the electron mass and coupling constant appearing in the Lagrangian of Quantum Electrodynamics.

In effective theories the values of parameters are assumed to be known. E.g. when analyzing the problem of the hydrogen atom spectrum using the Schrodinger equation, the electron and proton mass and the value of the fine structure constant are given; it is assumed that these values will be established in some  ``high-brow'', fundamental theory. Quantum Field Theory is the most promising candidate for this task.  Of course, it is impossible to {\it derive} values of the parameters (like charge and mass) in the framework of Quantum Field Theory alone - some numeric values, serving as an input data are necessary. Then, the best bet in establishing the values of the parameters would be the extraction of their values by comparing results following from the Quantum Field Theory with these input data. This procedure is described in the present note on the example of Quantum Electrodynamics the Lagrangian of which contains two parameters - fermion mass $m$ and the coupling constant $e$.



We will concentrate on $e$, assuming that the value of parameter $m$ is known. Establishing the numeric value of $m$ is a slightly easier task from theoretical viewpoint since mass is the invariant of Poincare group and does not require for its definition interacting fields, while coupling constant can not be introduced for a free fields - at least first non-trivial order of perturbation theory or some other approximation to an interacting quantum fields description is required.

To proceed with the comparison of the results of Quantum Field Theory with the experimental data we need to make sure that the field-theoretical expressions contain no divergences. Mathematical method used to demonstrate that the results of calculations in Quantum Field Theory may be finite is called renormalization. Renormalization procedure is sketched below.
\subsection*
{\bf Brief description of renormalization}
As is well known, the available methods of calculation in the framework of Quantum Field Theory encounter infinities that either persist, leading to nonsensical results, or cancel among themselves, leaving finite parts \cite{bjerken}. In order to operate with the well defined quantities the regularization is introduced. The idea is to distort the original field-theoretical expressions so that mathematical manipulations become well defined. At the end of calculations the distortion parameter is set to the initial value. 

As an example, consider integral in the four dimensional space-time:
\[
J(1,\,4)\,=\,\int\,{d^{4}k\over (k^{2}-m^{2})^{1}}
\]
In the regime $|k|\,\gg\,m$ we have $J(1,\,4)\approx \int\,d^{4}k/k^{2}$ which leads to an infinite result when integrating over all space.

There are many ways to regularize divergences. In case of integral $J(1,\,4)$ 
one can introduce cut-off $|k|\,\leq\,\Lambda$, or change the power of $(k^{2}-m^{2})$ from $-1$ to $-s$, or change the dimensionality of space time from $4$ to any number $n$. Correspondingly, instead of $J(1,\,4)$ we will have:
\[
J(1,\,4;\,\Lambda)\,=\,\int^{k^{2}<\Lambda^{2}}\,{d^{4}k\over k^{2}-m^{2}},\quad J(s,\,4)\,=\,\int\,{d^{4}k\over (k^{2}-m^{2})^{s}},\quad J(1,\,n)\,=\,\int\,{d^{n}k\over k^{2}-m^{2}}
\]
When $\Lambda$ is finite, or when $s\,>\,2$, or when $n\,<\,2$ integrals do not diverge and the mathematical operations involving these integrals are well defined. After the desired result is expressed in terms of regularized quantities (in our case, in terms of $J(4,\,1;\,\Lambda)$, or $J(4,\,s)$, or $J(n,\,1)$), the ``distortion parameter'' is set back to its initial value: $\Lambda=\infty$, or $s= 1$ or $n=4$. 

In Quantum Electrodynamics (QED), containing only two input parameters - coupling constant $e$ and the electron mass $m$, prescription leading to a  finite results can be described as follows \cite{dyson}: use regularization (say, dimensional regularization, prescribing to use $n$ dimensional space-time at the intermediate stage of calculations \cite{dimensional})  to calculate any two physical quantities $\Sigma_{\alpha}$ and $\Sigma_{\beta}$:
\begin{equation}
\label{hysquantities}
\Sigma_{\alpha}=\Sigma_{\alpha}(e_{0},m_{0};n),\quad\Sigma_{\beta}=
\Sigma_{\beta}(e_{0},m_{0};n)
\end{equation}
where $e_{0}$ and $m_{0}$ are the input, unrenormalized coupling constant and the electron mass and the momentum dependence, being irrelevant for the discussion,  is omitted. Standard choice for $\Sigma_{\alpha}$ and $\Sigma_{\beta}$ are the renormalized mass and coupling constant \cite{bjerken}:
\begin{equation}
\label{standard}
\Sigma_{\alpha}\,=\,m\,=\,m_{0}\,(1\,+\,e^{2}_{0}\,Z_{m}^{(1)}\,+\,e^{4}_{0}Z_{m}^{(2)}\,+\cdots),\;
\Sigma_{\beta}\,=\,e\,=\,e_{0}\,(1\,+\,e^{2}_{0}\,Z_{e1}^{(1)}\,+\,e^{4}_{0}Z_{e2}^{(2)}\,+\cdots),
\end{equation}
but, in general, any two physical quantities can be used.

In most cases, when the regularization is removed, expressions for $\Sigma_{\alpha}$ and $\Sigma_{\beta}$ diverge. Therefore, regularization can not be removed, i.e. the limit $n\rightarrow 4$ does not exist in (\ref{hysquantities}). Since $\Sigma_{\alpha}$ and $\Sigma_{\beta}$ are physical quantities, they {\it must} be finite\footnote{Definition of physical quantities of Quantum Field Theory is a highly non-trivial problem; e.g. contrary to what may seem self-evident,  the matrix element of elastic electron-electron scattering diverges - only the matrix element accounting for the emission of infrared photons are finite, and thus can be interpreted as a physical quantities, representing a measurable \cite{bjerken}.}. 
The ansatz  implicitly embedded in renormalization procedure is that $\Sigma_{\alpha}$ and $\Sigma_{\beta}$ are finite and it is just the imperfect math  which prevents to demonstrate the finiteness of physical quantities. 

Theory is called renormalizable if there are finite number of physical quantities $\Sigma_{j},\;j=1,\,2\cdots N$ for which regularization can not be removed and the remaining $\Sigma_{N},\,\Sigma_{N+1},\,\cdots$ do not diverge after the regularization is  removed. In other words, in renormalizable theory calculation of $\Sigma_{N},\,\Sigma_{N+1},\,\cdots$ leads to an explicitly finite expression. Theory is called non renormalizable if there is no physical quantity which remains finite after the regularization is removed\cite{bjerken}. 

Now, back to QED. It is renormalizable theory as it was demonstrated more than half a century ago \cite{dyson}.

Renormalization procedure for QED prescribes to solve equations (\ref{hysquantities}) for $e_{0}$ and $m_{0}$ in terms 
of $\Sigma_{\alpha}$, $\Sigma_{\beta}$ and the regularization parameter $n$, and substitute into the expression for any 
other physical quantity $\Sigma_{\gamma}$. In QED expression for $\mit\Sigma_{\gamma}$ becomes finite when the regularization is removed,  i.e. the limit $n\rightarrow 4$ exists:
\begin{eqnarray}
\label{sigma3}
\sigma_{\gamma}\,\equiv\,\lim_{n\rightarrow 4}\Sigma_{\gamma}(e_{0},m_{0};n)=
\lim_{n\rightarrow 4}\Sigma_{\gamma}(e_{0}(\Sigma_{\alpha},\Sigma_{\beta};n),
m_{0}(\Sigma_{\alpha},\Sigma_{\beta};n);n)
\,<\,\infty
\end{eqnarray}
After the divergences are removed it becomes possible to compare the expression for a physical quantity with the numeric value for that quantity extracted from the experiment. This comparison allows to establish the numeric value of renormalized parameters.  E.g. the numeric value of renormalized coupling constant $e$ is extracted according to following scheme:
\begin{equation}
\label{sigma4}
\sigma(m,\,e)_{th}\,=\,\sigma_{exp}\,\rightarrow\,e\,=\,f(\sigma_{exp})
\end{equation}
where $\sigma_{exp}$ is the measured value of physical quantity $\sigma$ and $\sigma_{th}$ is the expression for $\sigma$ obtained in the framework of QED. This procedure is described below. 

\subsection*{Extracting value of $e$ by comparing QED calculations with the Coulomb law}

As an observable $\sigma$ let us use the field created by a static source with the mass $M\,\gg\,m$ where $m$ is the electron mass. The proton at rest can be chosen as such a source. From the experiment it is known that at large distances, $r\,\gg\,\hbar/mc$, the electromagnetic field generated by a proton is:
\begin{equation}
\label{pot}
A_{0}(t,\,{\bf r})\,=\,{Q\over r},\;\;{\bf A}(t,\,{\bf r})\,\approx\,0
\end{equation}
where
\begin{equation}
\label{q}
Q^{2}\approx {4\pi\over 137}
\end{equation}

Equation (\ref{q}) can not be derived from QED; using (\ref{q}) as an input, we have to calculate the value of $e$ from QED.

We need to calculate the long distance asymptotic of electromagnetic  field $A_{\mu}$ created by a static source, compare the result  expressed in terms of $e$ and $m$ to the experimental information (\ref{pot})-(\ref{q}), and from that comparison extract the numeric value for $
e$.

From QED it follows that field created by the static source $j_{\mu}=e\delta_{\mu 0}\delta({\bf r})$
is
\begin{equation}
\label{ph}  
A_{0}({\bf r})\,=\,\int\,D^{0\nu}(r-r^{\prime})\,j_{\nu}(r^{\prime})\,d^{4}r^{\prime}\,=\,e\,\int\,dt\,D({\bf r}^{\prime})\,=\,e\,\int\,d^{3}k\,e^{i{\bf kr}}\,D(0,{\bf k})
\end{equation}
In (\ref{ph}) $D\equiv D^{00}$ is a scalar part of the photon propagator $D^{\mu\nu}$ and $e$ is the QED coupling constant.

In a first non-trivial order of perturbation theory expression for $D$ is (from now on we omit inessential overall factors and use units $\hbar=1,\;c=1$):
\begin{eqnarray}
\label{D}
D({\bf k})&=&{1\over {\bf k}^{2}}\Biggl\{1+{2\alpha\over \pi}\Biggl[\int^{1}_{0}\,dx\,x(1-x)\Biggl(\ln{m^{2}-\mu^{2}x(1-x)\over m^{2}+{\bf k}^{2}x(1-x)}\,+\cr\cr\cr
&+&{\mu^{2}\over {\bf k}^{2}}
\,{({\bf k}^{2}+\mu^{2})x(1-x)\over m^{2}-\mu^{2}x(1-x)}\Biggr)\Biggr]+{\cal O}(\alpha^{2})\Biggr\}
\end{eqnarray}
where
\begin{equation}
\label{alp}
\alpha\,\equiv\,{e^{2}\over 4\pi}
\end{equation}
and  $e\,=\,e_{0}\,+{\cal O}(e^{3}_{0})$ is a renormalized coupling constant. 

(\ref{D}) follows from the standard perturbation theory \cite{bjerken} and the straightforward algebraic manipulations; for our goal it is enough just to sketch the derivation and explain what is $\mu$ and how did it appear in (\ref{D}).

Calculation of the photon propagator requires the knowledge of the vacuum polarization operator $\Pi_{\mu\nu}$: relation between these two is \cite{bjerken}
\begin{equation}
\label{DP}
\Biggl[k^{2}\,g_{\mu\lambda}\,-\,e^{2}_{0}\,\Pi_{\mu\lambda}(k)\Biggr]\,D_{\lambda\nu}(k)\,=\,-\,g_{\mu\nu}
\end{equation}
Photon propagator is obtained by inverting equation (\ref{DP}). 
Polarization operator diverges; in order to obtain finite result for a photon propagator it is necessary and sufficient to get rid of infinities in $\Pi(k^{2})$ where $\Pi(k^{2})$ is a form-factor of the polarization operator: $\Pi_{\mu\nu}(k)\,=\,(k^{2}\,g_{\mu\nu}\,-\,k_{\mu}k_{\nu})\,\Pi(k^{2})$. This may be achieved by subtracting twice at some normalization point $k^{2}\,=\,\mu^{2}$:       
\begin{eqnarray}
\label{pi2}
\Pi_{R}(k^{2},\,\mu^{2}) = \Pi(k^{2})\,-\,\Pi(\mu^{2})\,-\,(k^{2}-\mu^{2})\,\Pi^{\prime}(\mu^{2});\quad\quad \Pi_{R}(\mu^{2},\,\mu^{2}) = 0
\end{eqnarray}
That is how the (arbitrary) normalization scale $\mu$ appears in $\Pi$ and through (\ref{DP}) appears in the expression for the photon propagator (\ref{D}). Exact expression (in case of perturbation theory - summed up series) for a physical quantity does not depend on an unphysical parameter $\mu$ \cite{bjerken}; since we use the first order approximation, (\ref{D}) contains $\mu$ explicitly.

Standard choice for the subtraction point in QED is $\mu^{2}=0$.  Expression for the photon propagator appearing in a textbooks \cite{bjerken} is obtained using $\mu^{2}=0$; for our purposes we have to consider the case of arbitrary $\mu\neq 0$ and then find out which values of $\mu$ allow to calculate long-range asymptotic of $A_{0}$ in QED.

Now, since we need asymptotic at large $r$, let us expand $D$ in ${\bf k}^{2}$:
\[
D({\bf k})\equiv{1\over {\bf k}^{2}}\Biggl\{1\,+\,{2\alpha\over \pi}\Biggl[\int^{1}_{0}\,dx\,x(1-x)\,\Biggl(\ln\biggl( 1-{\mu^{2}x(1-x)\over m^{2}}\biggr)\,+\,{\mu^{2}x(1-x)\over m^{2}-\mu^{2}x(1-x)}\,+
\]
\begin{eqnarray}
\label{D1}
+\,{\mu^{4}\,x(1-x)\over {\bf k}^{2}\,(m^{2}-\mu^{2}x(1-x))}-{\bf k}^{2}\,{x(1-x)\over m^{2}}\,+\,{\cal O}({\bf k}^{4})\Biggr)\Biggr]\,+\,{\cal O}(\alpha^{2})\Biggr\}
\end{eqnarray}

First two terms in $2\alpha[\;\;]/\pi$ of (\ref{D1}),
\[
\int^{1}_{0}\,dx\,x(1-x)\,\Biggl(\ln\biggl( 1-{\mu^{2}x(1-x)\over m^{2}}\biggr)\,+\,{\mu^{2}x(1-x)\over m^{2}-\mu^{2}x(1-x)}\,\Biggr)
\]
are ${\bf k}$-independent; we denote the corresponding Fourier transform by
\begin{equation}
\label{f}
{2\,\alpha\, F(m,\,\mu)\over \pi\,r}
\end{equation}
Third term of $2\alpha[\;\;]/\pi$ is $\sim {\bf k}^{-2}$; denote Fourier transform by
\begin{equation}
\label{f2}
\mu^{4}\,\varphi(m,\,\mu)\,r
\end{equation}
Fourier transform of the fourth term, $\sim {\bf k}^{2}$, results into
\begin{equation}
\label{f3}
-\,{\alpha \over 15\pi}\,\delta({\bf r})
\end{equation}
Coefficient in (\ref{f3}) is $\mu$ independent which means that the corresponding term should be directly interpreted as an observable. This is really the case: (\ref{f3}), known as the Uehling contact term \cite{uh} is observed in spectroscopical experiments.

Collecting these together in (\ref{ph}) results in
\begin{equation}
\label{ph1}
A_{0}({\bf r})\,=\,e\,\Biggl[{1\over r}\,+\,{e^{2}\over 4\pi}\,\Biggl({2F(m,\,\mu)\over \pi r}\,-\,{1\over 15\pi m^{2}}\delta({\bf r})\,+\,2{\mu^{4}\varphi(m,\,\mu\over \pi}\,r\Biggr)\,+\,{\cal O}(\alpha^{2})\Biggr]\,\equiv\,\,{Q(\alpha,\,m,\,\mu,\,r)\over r}
\end{equation}
Comparing the QED result (\ref{ph1}) with the experimental information (\ref{pot}) demonstrates that $Q(\alpha,\,m,\,\mu,\,r)$ is renormalization invariant and thus can be compared with $Q$, the coefficient of $1/r$ in the Coulomb law.
Note that so far the value of $e$ is not known; we need to simplify (\ref{ph1}) as much as possible and then compare the resulting expression to (\ref{pot})-(\ref{q}).



Expression (\ref{ph1}) contains term proportional to $r$ - $2\alpha\mu^{4}\varphi(m,\,\mu)\,r/\pi$. Therefore, in calculating asymptotic for large $r$, terms ${\cal O}(\alpha^{2})$ can not be neglected: it becomes necessary to sum up all the powers of $\alpha$, 
equate the result of the summation to $\sqrt{4\pi/137}$, and solve from this equation for the expansion parameter $\alpha=e^{2}/4\pi$. In other words, if the terms proportional to $r$ are present in expression given by a fixed order of perturbation theory, it is necessary to sum up all the terms in all orders of perturbation theory and obtain expression for a physical parameter $Q(\alpha,\,m,\,\mu,\,r)$ in a closed analytical form. Result will contain all the powers of $\alpha$ which means that the numeric value of the expansion parameter $\alpha$ can not be extracted from the numeric value of $Q(\alpha,\,m,\,\mu,\,r)$ - it is a highly non-trivial task to solve $\alpha$ from 
 $Q(\alpha,\,m,\,\mu,\,r)\,=\,\sum^{\infty}_{j=1}\,\alpha^{j}\,Q_{j}(m,\,\mu,\,r)$.

In QED there is a way out of this difficulty: the condition $m^{2}\neq 0$\footnote{Examining expression for the photon propagator (\ref{D}) demonstrates that $D$ diverges when $\mu^{2}=0$ {\it and} when $m^{2}=0$. This problem of infrared divergences  can be resolved in massless electrodynamics, but, it presents an unanswered challenge to Quantum Chromodynamics - theory of interacting quarks and massless gluons \cite{qcd}.} allows to subtract at $\mu^{2}\,=\,0$. When $\mu^{2}=0$ the term $\sim r$ drops out, and from QED it follows that in the limit $r\rightarrow \infty$ and when $\mu^{2}=0$  field is 
\begin{equation}
\label{ph4}
A_{0}({\bf r})|_{\mu^{2}=0,\,r\gg 1/m}\,=\,{e\over r}\,\Biggl[1\,+\,{e^{2}\over 4\pi}\,{2F(m,\,0)\over \pi}\,+\,{\cal O}(e^{4})\Biggr]
\end{equation}
Here the renormalized coupling constant is defined at $\mu^{2}=0$.  


Comparing the  theoretical expression (\ref{ph4}) with the experimental data (\ref{pot})-(\ref{q}) allows to establish the numerical value of the expansion parameter $\alpha$. In the first order of perturbation theory, i.e. retaining only $e$-term and neglecting $e^{3}$-term in (\ref{ph4}), we obtain
\begin{equation}
\label{aa1}
A_{0}({\bf r})\,=\,{Q\over r}\,\approx {e\over r}\;\rightarrow\;e\,=\,Q\;\rightarrow\;\alpha|_{\mu^{2}=0}\,=\,{e^{2}\over 4\pi}|_{\mu^{2}=0}\,\approx\,{Q^{2}\over 4\pi}\,=\,{1\over 137}
\end{equation}
The value 
\begin{equation}
\label{e}
{e^{2}\over 4\pi}|_{\mu^{2}=0}\,\approx\,{1\over 137}\,<\,1
\end{equation}
justifies assumption $e\,<\,1$ used in deriving expression (\ref{aa1}), when we have neglected term $\sim\,e^{3}$. If more accuracy is desired, then $e^{3}$-term is retained and the value of $e$ emerges as a solution of a cubic equation.

After the value of the expansion parameter is established, it is possible to find its value for a different subtraction point $\mu$. This is done by analyzing  the renormalization group equation \cite{rg}:
\begin{equation}
\label{em}
\mu\,{\partial e(\mu)\over \partial \mu}\,=\,{e^{3}(\mu)\over 12\pi^{2}}\end{equation}
To solve equation (\ref{em}), the boundary condition is necessary, and this boundary condition is not given by the renormalization group. This boundary condition, numeric value of renormalized coupling constant
\[
{e^{2}(0)\over 4\pi}\,=\,{1\over 137}
\]
is established from the comparison of the theoretical expressions with the experimental data as it is described above.


\subsection*{Discussion}
Let us summarize the procedure of establishing the numeric value QED coupling constant. Prescription can be described as follows:
\begin{itemize}
  \item Choose physical quantity $\sigma$. Numeric value of measurable $\sigma$ is known from the experiment.
  \item Calculate $\sigma$ in the framework of renormalized Quantum Field Theory. In QED with the two input parameters the expression is $\sigma(m,\,e)_{th}$.
  \item Provided that the value of $m$ is known, find $e$ from the equation
  \[
  \sigma(m,\,e)_{th}=\sigma
  \]
\end{itemize}
For $\sigma$ we have chosen $Q$, the coefficient of $1/r$ appearing in the Coulomb law. As it is mentioned earlier, any physical quantity will do fine. E.g., one can determine numeric value of $e$ from the measurement of anomalous magnetic moment of electron \cite{mm} or quantized Hall conductivity \cite{hall}. Choosing Coulomb law seems bit more intuitive and transparent since there is no need to involve measurable depending on purely quantum features of electrodynamics (anomalous magnetic moment) or to apply QED to a problems of mesoscopic physics (Quantum Hall Effect). Of course, any choice for $\sigma$ yields the same result as in (\ref{e}), just the accuracy may be different. Establishing the value of the QED coupling constant with the best possible accuracy was not our aim here - this goal requires some ``heavy-weight'' calculations, e.g., calculation of the anomalous magnetic moment of the muon in higher orders of perturbation theory, accounting for all the degrees of freedom of the Standard Model (electron, muon and tau leptons, quark, and $W$ and $Z$ bosons) \cite{hagi}. 

\end{document}